%% file: arxiv.tex
\documentclass[floatfix,aps,prl,superscriptaddress,reprint]{revtex4-2}
\usepackage{graphicx}
\usepackage{amsmath}
\usepackage{amssymb}
\usepackage{nameref}
\usepackage{hyperref}
\usepackage{cleveref}

\begin{document}

\title{Origin of circular and triangular pores in electron-irradiated hexagonal boron nitride}

\author{Umair Javed}
\email{umair.javed@univie.ac.at}
\affiliation{University of Vienna, Faculty of Physics, Boltzmanngasse 5, 1090 Vienna, Austria}
\affiliation{University of Vienna, Vienna Doctoral School in Physics, Boltzmanngasse 5, 1090 Vienna, Austria}

\author{Manuel L\"angle}
\affiliation{University of Vienna, Faculty of Physics, Boltzmanngasse 5, 1090 Vienna, Austria}

\author{Vladimír Zobač}
\affiliation{University of Vienna, Faculty of Physics, Boltzmanngasse 5, 1090 Vienna, Austria}

\author{Alexander Markevich}
\affiliation{University of Vienna, Faculty of Physics, Boltzmanngasse 5, 1090 Vienna, Austria}

\author{Barbara Maria Mayer}
\affiliation{University of Vienna, Faculty of Physics, Boltzmanngasse 5, 1090 Vienna, Austria}
\affiliation{Uppsala University, Faculty of Physics, Box 516, SE-751 20, Sweden}

\author{Clara Kofler}
\affiliation{University of Vienna, Faculty of Physics, Boltzmanngasse 5, 1090 Vienna, Austria}
\affiliation{University of Vienna, Vienna Doctoral School in Physics, Boltzmanngasse 5, 1090 Vienna, Austria}

\author{Martin Paul}
\affiliation{University of Vienna, Faculty of Physics, Boltzmanngasse 5, 1090 Vienna, Austria}

\author{Darwin Lorber}
\affiliation{University of Vienna, Faculty of Physics, Boltzmanngasse 5, 1090 Vienna, Austria}

\author{Nandhini Ravindran}
\affiliation{University of Vienna, Faculty of Physics, Boltzmanngasse 5, 1090 Vienna, Austria}
\affiliation{University of Vienna, Vienna Doctoral School in Physics, Boltzmanngasse 5, 1090 Vienna, Austria}

\author{Clemens Mangler}
\affiliation{University of Vienna, Faculty of Physics, Boltzmanngasse 5, 1090 Vienna, Austria}

\author{Toma Susi}
\affiliation{University of Vienna, Faculty of Physics, Boltzmanngasse 5, 1090 Vienna, Austria}

\author{Jani Kotakoski}
\email{jani.kotakoski@univie.ac.at}
\affiliation{University of Vienna, Faculty of Physics, Boltzmanngasse 5, 1090 Vienna, Austria}

\begin{abstract}
For nearly two decades, it has been known that electron irradiation of hexagonal boron nitride (hBN) in a transmission electron microscope leads to the formation of triangular pores.
This has been attributed to the lower displacement threshold energy of boron, with or without the assistance of an inelastic scattering event, typically assuming that chemical processes caused by residual gases can be neglected.
In this study, in contrast to previous high-vacuum experiments, we show that electron irradiation in ultra-high vacuum leads to circular pores, whereas even small amounts of oxygen in the atmosphere during the experiment drive the pores to grow into triangle shapes with nitrogen-terminated edges.
This result is shown to hold for samples of different types and from different manufacturers, and at different electron energies as well as focused scanning and defocused stationary beams.
Our results explain the chemical origin of triangular pores in hBN and demonstrate a deterministic way to create atomically defined pores in this important 2D material.
\end{abstract}

\maketitle

\section{Introduction}
Two-dimensional~(2D) hexagonal boron nitride~(hBN), a wide-bandgap semiconductor with a honeycomb structure of alternating B and N atoms, is an excellent candidate for many applications, such as an electrocatalyst for oxygen evolution reaction~\cite{khan_2d_2017}, as host of quantum emitters~\cite{tran_quantum_2016}, as membrane for DNA sequencing~\cite{zhou_dna_2013}, and as an active component in energy conversion and storage technologies~\cite{han_functionalised_2020}.
Many of the potential applications depend on defect engineering, including creating pores in the material, and a significant effort has been made to understand both defect formation and pore growth in hBN~\cite{weng_functionalized_2016,byrne_fabrication_2026}.
Transmission electron microscopy (TEM) is an important tool for creating and examining defects in 2D materials~\cite{su_nanopores_2021,thiruraman_angstrom-size_2018}, where an extended exposure to the electron beam can lead to pore formation~\cite{fischbein_electron_2008}.
Unlike in an excellent conductor such as graphene~\cite{susi_isotope_2016, meyer_accurate_2012}, in relative insulators such as hBN, electron irradiation damage is a combination of elastic and inelastic interactions~\cite{susi_quantifying_2019, bui_creation_2023}.
Additionally, it has been shown that even a partial pressure of oxygen as low as $10^{-8}$~mbar in the microscope column contributes to the defect formation and pore growth in 2D materials~\cite{leuthner2019scanning,leuthner2021chemistry,aahlgren2022atomic}.
However, the situation for hBN remains unclear.

Formation of pores in hBN under electron irradiation is of particular interest due to their characteristic triangular shape, known for nearly two decades~\cite{jin2009fabrication,ryu2015atomic,meyer2009selective,alem2009atomically,byrne2025atomic}.
The broadly accepted explanation for this was that boron is easier to displace than nitrogen, owing to their different displacement cross sections resulting from their mass difference and differing displacement threshold energies~\cite{kotakoski2010electron}.
However, these values correspond to knock-on damage only assuming an elastic scattering event, which has been shown to not fully describe electron beam damage in 2D materials that are not as good conductors as graphene~\cite{kretschmer_formation_2020, speckmann_combined_2023, bui_creation_2023}. Surprisingly, the most precise recently measured cross-sections for B and N ejection under both 60 and 80-keV irradiation are equal within the uncertainty of the experiment~\cite{bui_creation_2023}.
Both nitrogen and boron single vacancies~\cite{bui_creation_2023}, and even boron-terminated tetravacancies~\cite{byrne_fabrication_2026,cretu_structure_2015}, have been reported to form under electron irradiation at room temperature.

In these experiments, all larger structures have corresponded to triangular shapes with nitrogen-terminated edges.
Non-triangular pores have only been reported when the electron beam was condensed to increase the current density to get hexagonal pores~\cite{gilbert_fabrication_2017}, or as a result of ``drilling'' with a focused 200-keV electron beam onto a few-layer hBN sample to get round pores~\cite{keneipp_nanoscale_2024}.
However, at elevated temperatures, the picture is more complicated.
For example, Cretu {\it et al.}~\cite{cretu_inelastic_2015} demonstrated that at 1200$^\circ$C hexagonal pores form, whereas at 800$^\circ$C they are circular.
However, in that study, the pores were triangular at room temperature.
While it has been speculated that the residual vacuum in the microscope could also play a role in the preferred removal of boron atoms~\cite{gilbert_fabrication_2017,cretu_inelastic_2015, kotakoski2010electron}, this hypothesis has not been tested due to the required customized experimental equipment.

Here, extending our earlier work with graphene~\cite{leuthner2019scanning,leuthner2021chemistry}, MoS$_2$ and MoTe$_2$~\cite{aahlgren2022atomic}, we systematically study pore growth in monolayer hBN in ultra-high vacuum (UHV) and in low-pressure oxygen atmospheres under electron irradiation at room temperature.
We show that even a slight amount of oxygen leads to increased pore growth compared to UHV.
We also show that whereas pores created in UHV are circular, they become triangular when as little as $10^{-9}$~mbar of oxygen is introduced, well below the residual vacuum pressure in typical TEM columns.
These results are confirmed across samples of different types and sources, diverse illumination modes, and different electron energies.
Besides explaining the formation of triangular pores reported in the literature, our results suggest a route towards atomically precise engineering of pore shapes in hBN.
Pore shape directly influences the edge chemistry, and local electronic and transport properties~\cite{vicarelli2015controlling}.
Triangular pores provide anisotropic, while circular pores provide more isotropic interaction landscapes, which can affect DNA translocation and signal stability during sequencing~\cite{feng2015identification}, as well as selectivity and permeability in filtration~\cite{surwade2015water}.

\section{Results and discussion}
Commercially available hBN grown via chemical vapor deposition (CVD) from Graphene Supermarket (CVD1) and Sigma-Aldrich (CVD2), as well as mechanically exfoliated hBN from HQ Graphene, was transferred onto Quantifoil\textsuperscript{\texttrademark} TEM Au grids with 2~µm diameter holes (see Methods).
Unless otherwise mentioned, the experiments were carried out with CVD1 samples.
Prior to insertion into the microscope, the samples were baked at 150°C for about 12~h in vacuum.
The modified Nion UltraSTEM 100 in Vienna~\cite{hotz_ultra-high_2016}, an aberration-corrected scanning transmission electron microscopy (STEM) instrument, was used for the experiments.
Images were recorded with the medium-angle annular dark-field (MAADF) detector.
The base pressure in the sample chamber is in the order of $10^{-10}$~mbar, and atomic-resolution imaging can be maintained to up to a pressure of $4\times 10^{-6}$~mbar~\cite{leuthner2019scanning}.
Acceleration voltages of 60~kV and 80~kV were used for imaging.

\begin{figure*}
    \centering
    \includegraphics{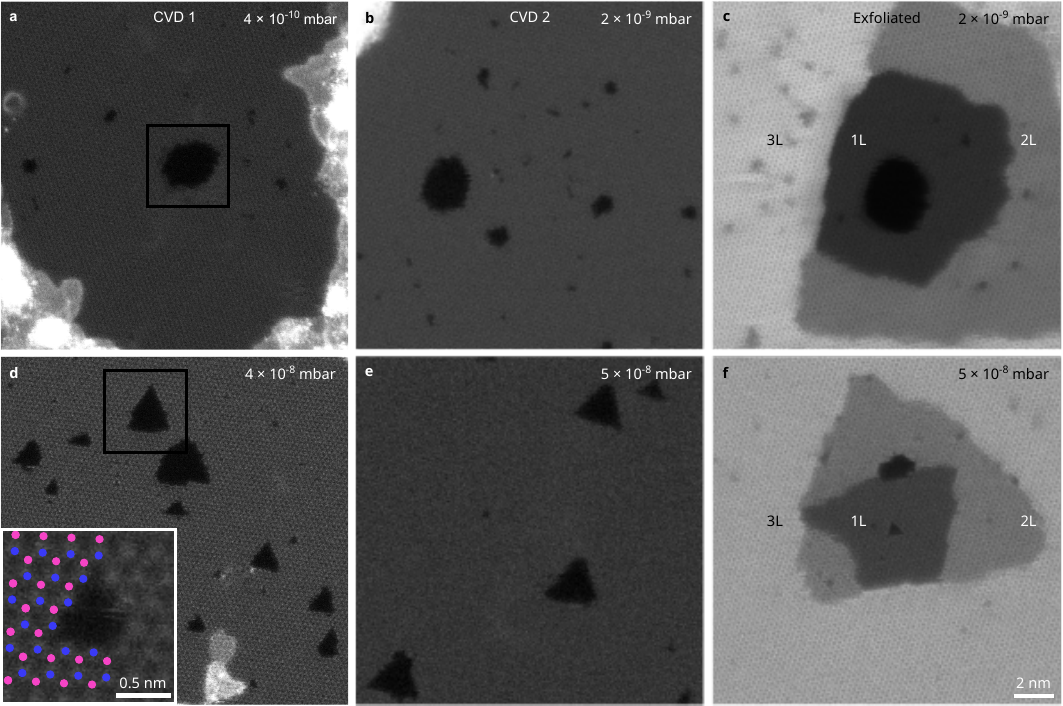}
    \caption{{\bf Circular and triangular pores in different hBN samples.} STEM-MAADF images of pores formed in (a) CVD1, (b) CVD2, and (c) exfoliated samples in ultra-high vacuum. STEM-MAADF images of pores formed in (d) CVD1, (e) CVD2, and (f) exfoliated samples under oxygen partial pressure of around~$10^{-8}$~mbar; the scale bar applies to all images. Inset in (d) shows a larger-magnification image of a pore with a partial overlay showing atomic sites (blue for nitrogen and pink for boron).}
    \label{fig: Figure 1}
\end{figure*}

The effect of prolonged imaging at UHV at 60~kV is shown in Fig.~\ref{fig: Figure 1}(a-c).
When hBN is imaged continuously in UHV, vacancy defects emerge~\cite{bui_creation_2023} and grow into pores.
However, no preference for either boron- or nitrogen-terminated edges is seen, and the pores appear round, in contrast to other reported room-temperature experiments~\cite{jin2009fabrication,meyer2009selective,ryu2015atomic,dai_evolution_2023,byrne2025atomic,yin_triangle_2010,cretu_inelastic_2015,gilbert_fabrication_2017,alem2009atomically,xu_nucleation_2022}.
However, when oxygen is introduced into the column, corresponding to a pressure in the order of $10^{-8}$~mbar at the sample, the appearing pores are triangular, as shown in Fig.~\ref{fig: Figure 1}(d-f).
The image with a smaller field of view, overlaid on Fig.~\ref{fig: Figure 1}d, shows a larger magnification of one of the pores in the same sample.
At this magnification, it is easy to distinguish the brighter nitrogen from the less bright boron atoms due to their MAADF contrast (the intensity corresponds approximately to $Z^{1.64}$~\cite{krivanek_atom-by-atom_2010}, where $Z$ is the atomic number), which makes it clear that the triangular pores are terminated by nitrogen.
The same holds for all triangular pores observed in this study, as also confirmed by the orientation of the pores with respect to the lattice~\cite{meyer2009selective}.
As can be seen in Fig.~\ref{fig: Figure 1}, the result holds for all three sample types (CVD-grown hBN from different manufacturers and mechanically exfoliated hBN), and therefore does not depend on the intrinsic sample quality, including its defect distribution and defect types, but is rather an intrinsic property of the material.

\begin{figure}[t]
    \centering
    \includegraphics{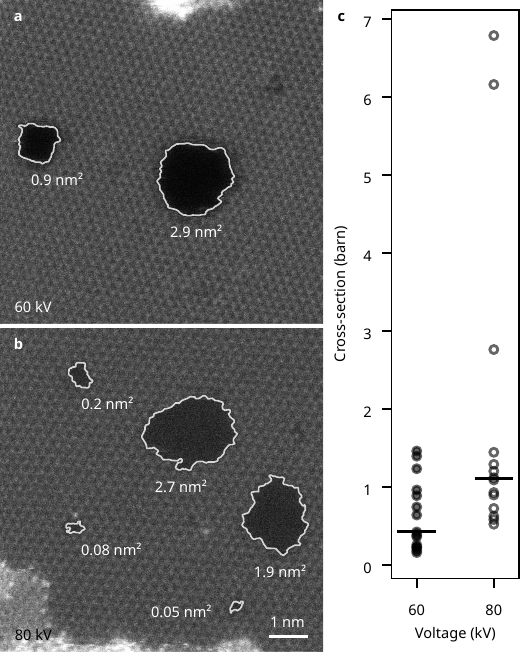}
    \caption{{\bf Pore growth at different electron acceleration voltages.} (a) STEM-MAADF image of round pores in hBN imaged at 60 kV. The pores are highlighted using semi-automatic image analysis. The white lines show the approximate circumferences of each pore, with their areas denoted below. (b) STEM-MAADF image of pores recorded at 80~kV. The scale bar applies to both images. (c) Displacement cross-section of edge atoms at 60 and 80~kV (with filled and open symbols, respectively). Each symbol corresponds to one image sequence, and the horizontal line segment denotes the median of each data set.}
    \label{fig: Figure 2}
\end{figure}

In order to minimize unwanted damage~\cite{bui_creation_2023}, most of our experiments were carried out with an acceleration voltage of 60~kV.
Since this is below the energies used in most studies in the literature~\cite{jin2009fabrication,meyer2009selective,kotakoski2010electron,ryu2015atomic,dai_evolution_2023,byrne2025atomic,yin_triangle_2010,cretu_inelastic_2015,gilbert_fabrication_2017,alem2009atomically,xu_nucleation_2022}, we turn to the role of the electron energy.
A comparison of 60 and 80~keV in UHV is shown in Fig.~\ref{fig: Figure 2}.
We again observe only circular pores, confirming that this shape is not due to the low electron energy.
The circular shape indicates that nitrogen and boron atoms at the edge are removed at roughly the same rate and therefore must have similar cross-sections, as is the case for the pristine structure~\cite{bui_creation_2023}.
To quantify this effect, we next estimate the displacement cross-section for edge atoms.
The pore area ($A$) is measured semi-automatically via intensity thresholding, as can be seen in Fig.~\ref{fig: Figure 2}(a,b), where we show the pore areas and pore contours overlaid on the original images.
The number of atoms ejected ($N_\mathrm{ejected}$) in each frame was estimated as
\begin{equation}
    N_\mathrm{ejected} = \frac{\Delta A}{A_\mathrm{atom}},
    \label{eq: number of ejected atoms}
\end{equation}
where $\Delta A$ is the difference between areas of the pores in subsequent frames and $A_\mathrm{atom}$ is the area of an individual atom in hBN corresponding to half of the unit cell (0.0272~nm$^2$).
The number of edge atoms~$N_\mathrm{edge}$ was approximated as
\begin{equation}
    N_\mathrm{edge} = \frac{\sqrt{4\pi A}}{a},
    \label{eq: number of edge atoms}
\end{equation}
assuming a circular shape for the pores and that each atom occupies a length of $a$, the lattice parameter of hBN (0.251~nm), along the circumference. 
The displacement cross-section $\sigma$ for the edge atoms was then calculated as
\begin{equation}
    \sigma = \frac{N_\mathrm{ejected}}{N_\mathrm{edge}}\frac{1}{\phi},
\end{equation}
where $\phi$ is the electron fluence (number of electrons per unit area per frame).
To ensure that we measure only pore growth and not other effects, we include in the analysis only pores that are larger than 0.5~nm$^2$ at the beginning and do not merge with other pores during the experiment.
The estimated cross-sections are shown in Fig.~\ref {fig: Figure 2}c.
Despite the obvious uncertainties inherent to each estimation, the cross-section at 60~kV is observably lower than that at 80~kV, with the median values of 0.4~barn and 1.1~barn, respectively.
In comparison, for the pristine lattice, the cross-sections for boron atom ejection were reported as 0.023$\pm$0.006~barn (60~kV) and 0.09$\pm$0.02~barn (80~kV), while for nitrogen these were 0.018$\pm$0.005~barn and 0.08$\pm$0.01~barn, respectively~\cite{bui_creation_2023}.

\begin{figure*}[t]
    \centering
    \includegraphics{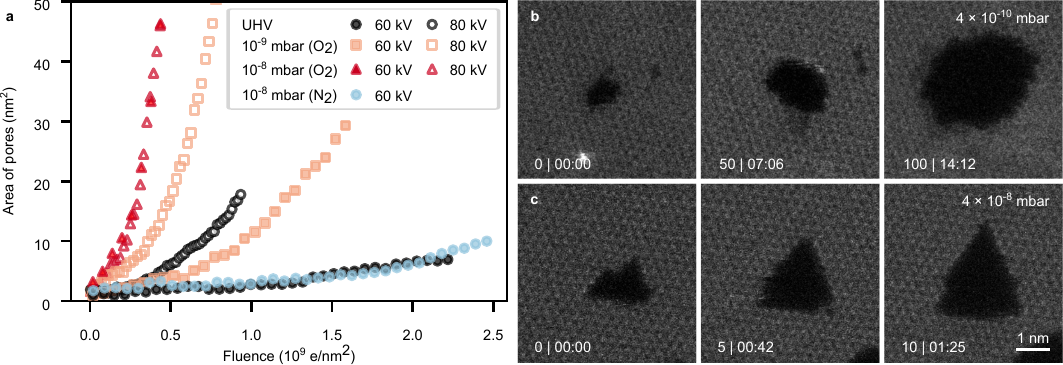}
    \caption{{\bf Pore growth under different atmospheres.} (a) Pore growth rate for different acceleration voltages and atmospheres. (b) A series of cropped STEM-MAADF images of pore growth at $4\times10^{-10}$~mbar (UHV). (c) A series of cropped STEM-MAADF images of pore growth at $4\times10^{-8}$~mbar oxygen. The uncropped images for the last images in both series are shown in Fig.~\ref{fig: Figure 1}(a\&d). The frame number and the total time (min:s) are noted on the bottom-left corners of the images. The scale bar applies to all images.} 
    \label{fig: Figure 3}
\end{figure*}

We then return to the initial observation that the sample atmosphere influences pore formation.
In the next experiments, the sample was exposed to different nitrogen and oxygen partial pressures, carefully introduced into the column during imaging, as described in Ref.~\cite{leuthner2019scanning}. 
We note that all pressure values given in this study correspond to the readings of the objective area gauge.
Pore growth was estimated by calculating the total pore area in each image using the method introduced in Fig.~\ref{fig: Figure 2}.
We note that pore growth can only occur after an initial point defect has been created, which, due to the stochastic nature of the process, occurs at different times and random locations.
Therefore, each imaged area contains pores of different sizes.
The results for different pressures and both acceleration voltages are shown in Fig.~\ref{fig: Figure 3}a.

There is a clear increase in pore growth rate with rising oxygen partial pressure.
In UHV at 80~kV, the rate of pore growth is higher than at 60~kV (see open vs. filled circles), as already discussed.
This is also true at an oxygen pressure of $10^{-9}$~mbar (open vs. filled rectangles), with the rate of pore growth rising at both voltages compared to UHV.
However, at $10^{-8}$~mbar (open vs. filled triangles), the pore growth rate at both voltages is similar, indicating that the amount of oxygen dominates the process.
To ensure that this phenomenon is indeed caused by oxygen gas in the column, we repeated the experiment in a nitrogen atmosphere (blue circles).
The results are comparable to those in UHV, demonstrating that nitrogen has no effect and confirming that oxygen is the primary reason for faster pore growth.

As expected from the results shown in Fig.~\ref{fig: Figure 1}, in addition to the faster pore growth rate, we also observe that the pore shape depends on the oxygen partial pressure.
As before, in UHV (Fig.~\ref{fig: Figure 3}b), the pores are round.
The pores remain round when the pressure is increased to $10^{-9}$~mbar, but at $10^{-8}$~mbar, pores again appear triangular with nitrogen-terminated edges~(Fig.~\ref{fig: Figure 3}c).
From these experiments, it is clear that even small amounts of oxygen in the microscope column significantly affect pore growth, altering the pore shape from round to triangular and increasing the growth rate.

\begin{figure*}[t]
    \centering
    \includegraphics{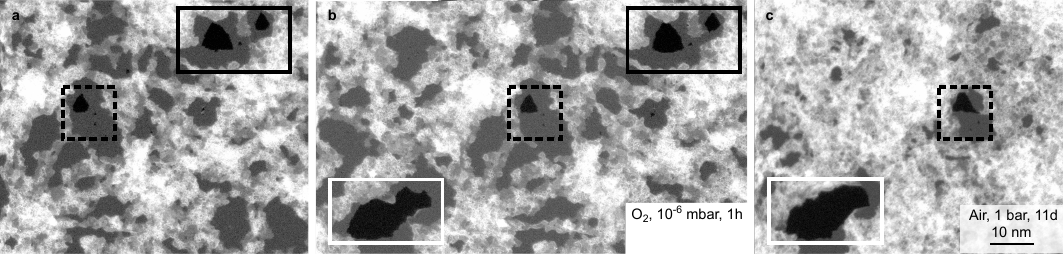}
    \caption{{\bf Influence of molecular oxygen.} MAADF-STEM images of (a) hBN with a few pores. (b) The same area after the sample was exposed to oxygen at $10^{-6}$~mbar pressure for 1~h, without the electron beam (composite of two images near the same field of view). (c) Same sample area after exposure to the ambient atmosphere for 11 days. The same pores are highlighted with corresponding rectangles. The scale bar applies to all images.}
    \label{fig: Figure 4}
\end{figure*}

To separate the effect of oxygen from that of the electron beam, the sample was also exposed to oxygen in the absence of the beam.
Fig.~\ref{fig: Figure 4}a shows a $64\times64$~nm² area containing a few pores.
After this image was captured, the beam was turned off, and oxygen was leaked into the column at a pressure of $10^{-6}$~mbar.
The sample was kept at that pressure for an hour, after which the leak valve was closed.
When the pressure dropped below $10^{-8}$~mbar, the sample was imaged again.
Fig.~\ref{fig: Figure 4}b shows that there are practically no changes in the sample due to mere oxygen exposure.
Next, the sample was taken out of the vacuum system and kept in an ambient atmosphere (laboratory air) for 11 days, with
Fig.~\ref{fig: Figure 4}c recorded afterward in UHV.
The surface contamination has changed because the sample was removed from the vacuum and heated after insertion to remove adsorbates; however, the smaller pores have remained practically unchanged, and only minor changes have occurred in the larger pores, presumably due to mechanical stress from sample handling.
However, it has previously been reported that air exposure can lead to a dramatic geometry change in hBN pores~\cite {dai_evolution_2023}. Our results suggest that such a geometry change may be due to an additional process and not the air exposure alone.

\begin{figure*}[t]
    \centering
    \includegraphics{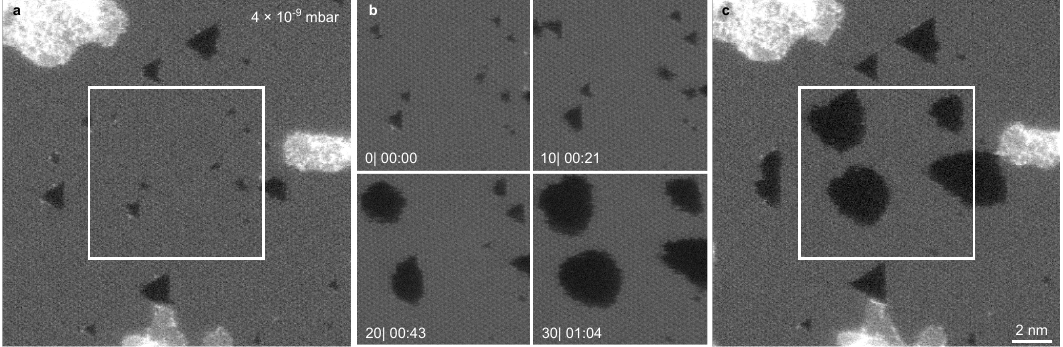}
    \caption{{\bf Pore-shape experiment at $4\times 10^{-9}$~mbar.} (a) A STEM-MAADF image at the beginning of the experiment.
    The white square shows the central area exposed to the electron beam during the image series shown in panel (b). Frame number and frame time are indicated in the bottom-left corners. (c) Image of the initial larger area recorded after the image series. The scale bar applies to all the images.}
    \label{fig: Figure 5}
\end{figure*}

Since the combination of the electron beam with the oxygen partial pressure in the column is responsible for triangular pores, we assume that molecular oxygen, split by the electron beam to atomic oxygen~\cite{leuthner2021chemistry}, is responsible for the observed changes.
To check whether this atomic oxygen leads to pore growth also without electron irradiation, we carried out further experiments at an oxygen partial pressure of $4\times10^{-9}$~mbar.
At this pressure, pores are usually round.
Fig.~\ref{fig: Figure 5}a shows an area where some pores have already been created by the beam.
In the next step, only the middle of the area, shown by the white square, was exposed to the electron beam.
Consecutive small frames in Fig.~\ref{fig: Figure 5}b show pore growth in that smaller area.
When the initial larger area is imaged again after approximately 1~min (Fig.~\ref{fig: Figure 5}c), we observe round pores in the irradiated smaller FOV.
However, outside that area, the pores remain triangular and somewhat larger than before.
This indicates that physical and chemical effects occur simultaneously, with physical effects dominating within the irradiated area, resulting in round pores. Outside this area, where minimal electron fluence (resulting from beam tails) is present but atomic oxygen is presumably available, chemical etching prevails.
This suggests that the observed damage in hBN can be attributed to a combination of physical damage induced by the electron beam and chemical damage caused by atomic oxygen.
Further, it is possible to determine the dominant type of damage from the pore shape. 

This observation also holds for irradiation with a defocused electron probe. Indeed, it has been shown that when converging the beam into a smaller area, pores become non-triangular where they were triangular before focusing~\cite{gilbert_fabrication_2017}.
To avoid potential effects from the scanning and focused electron probe and to adjust the local current density, we also conducted experiments with a defocused probe. 
In oxygen pressure of $4\times10^{-9}$~mbar and with a dose rate of $1.7\times10^{7}$ e$^{-}$s$^{-1}$nm$^{-2}$, we observe both triangular and round pores, showing that under these conditions the two mechanisms are equally prevalent.
However, when the pressure is increased to $1.3\times10^{-8}$~mbar while keeping the dose rate constant, the created pores are triangular.
At a lower dose rate of $1\times10^{6}$ e$^{-}$s$^{-1}$nm$^{-2}$, pores remain triangular at both partial presures of $4\times10^{-9}$ and $1.3\times10^{-8}$~mbar. 
These experiments are summarized in Supplementary Figs.~1--2, and Supplementary Table 1~\cite{supp2025}.
Together, these data allow us to estimate a quantitative correlation between the number of oxygen molecules and the number of electrons per area per time and the shape of the pores: when there are more than five oxygen molecules per ten million electrons, the pores that appear are triangular, whereas for fewer than three molecules, they are circular. In between these values, we find a mixture of shapes, including triangles, circles, and deformed hexagons.

To estimate the number of oxygen molecules at the surface at a given pressure, we used the kinetic theory of gases~\cite{ma_v_1860}, which is typically used for vacuum systems. Although the microscope column has additionally an electron beam that interacts locally with the atmosphere and the sample, it has a negligible influence on the composition of gases. 
The rate of impingement in vacuum~\cite{ohanlon2003} can be calculated by 
\begin{equation}
    J=\frac{P}{\sqrt{2\pi m k_\mathrm{B} T}},
\end{equation}
where $P$ is the pressure, $m$ is the molecular mass (in the case of the oxygen molecule, 32~amu), $k_\mathrm{B}$ is the Boltzmann constant, and $T$ is the temperature.
The impingement rate is multiplied by 2 for two surfaces in the monolayer (bottom and top), the frame time (8.5~s), and area to get the number of impinging molecules per frame ($J_{FOV}$), or with area per atom (0.0272~nm$^2$) for the number per atomic site ($J_{site}$).
For the estimate above, we have simply used the gauge pressure at the objective area to allow comparison across different instruments.
However, we previously estimated that, for the instrument used in these experiments, the pressure at the sample is about 10 times higher due to the specific geometry of the sample stage~\cite{leuthner2021chemistry}.
The impinging rate at different oxygen partial pressures used in this study, taking into account the additional factor of ten, is shown in Supplementary Table 2~\cite{supp2025}.

\begin{figure*}[t]
    \centering
    \includegraphics{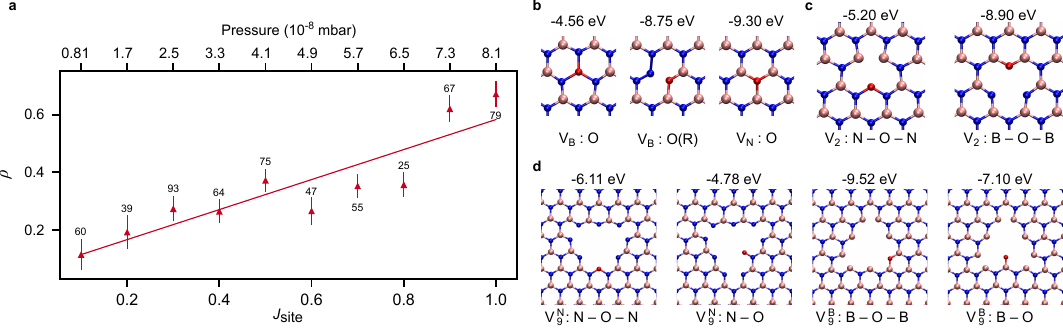}
    \caption{{\bf Oxygen absorption and edge-atom removal.} (a) Probability of removal of an edge atom ($\rho$) as a function of impinging O$_2$ molecules per atomic site per recorded image ($J_\mathrm{site}$), taking into account the additional factor of 10, as described in the text. The numbers next to the data points correspond to the number of individual values contributing to the mean. Atomic structures and DFT-calculated absorption energies of oxygen at different configurations of (b) single boron ($\mathrm{V_B}$) and nitrogen ($\mathrm{V_N}$) vacancies, (c) double vacancies ($\mathrm{V_2}$), and (d) nitrogen- ($\mathrm{V_9^N}$) and boron-terminated ($\mathrm{V_9^B}$) pores with nine missing atoms, along with the corresponding atomic structures. Pink spheres represent boron atoms, blue ones nitrogen, and red ones oxygen.}
    \label{fig: Figure 6}
\end{figure*}

The rate of pore growth shown in Fig.~\ref{fig: Figure 3}a is quadratic, since, as the pore becomes larger, the number of available atoms at the edges increases.
To evaluate in more detail the exact influence of oxygen partial pressure, we performed another set of experiments at 60~kV to minimize direct electron-beam damage.
In these experiments, the pressure is increased slowly, so that in every subsequent measurement, the impingement rate is increased by 0.1 molecules per atomic site per recorded image.
At each pressure, several pores were analyzed by estimating the probability of the removal of an edge atom ($\rho$) through linear regression from $N_\mathrm{ejected}$ as a function of $N_\mathrm{edge}$.
In Fig.~\ref{fig: Figure 6}a, $\rho$ is shown as a function of $J_\mathrm{site}$.
Each data point represents the linear regression comprising all measured pores at the given pressure, with error bars showing the corresponding standard errors of the means.
A linear fit to the data yields a slope of 0.52$\pm$0.06 removed edge atoms per O$_2$ molecule; {\it i.e.}, every fourth oxygen atom causes the removal of an atom from the edge.

Our experimental results show that oxygen radicals in the microscope column are responsible for the formation of triangular pores in hBN.
However, the mechanism by which boron atoms are preferentially removed and N-terminated edges form, remains unclear.
To gain further insight into the interaction of oxygen radicals with B and N atoms at the pore edges, we performed plane-wave density functional theory (DFT) simulations (Methods).
We find that on pristine hBN, an oxygen radical preferably adopts a bridge configuration between B and N atoms, which results in a significant elongation of the B--N bond by $\sim$9.6\%. 
Our calculated adsorption energy of -1.89\,eV is close to the previously reported value of -2.13\,eV~\cite{Zhao2012}.
Various configurations of oxygen radicals adsorbed on vacancy defects in hBN, along with their calculated adsorption energies, are shown in Fig.~\ref{fig: Figure 6}(b-d).
In particular, we considered single (V$_\text{B}$ and V$_\text{N}$, Fig.~\ref{fig: Figure 6}b) and double (V$_2$, Fig.~\ref{fig: Figure 6}c) vacancies, as well as nine-atom vacancies with either N- or B-terminated edges (V$_9^\text{N}$ and V$_9^\text{B}$, Fig.~\ref{fig: Figure 6}d) to represent small triangular pores.
The calculated adsorption energies range from -4.5 to -9.5~eV, indicating strong binding of O radicals to both nitrogen and boron edge atoms.
The adsorption energy of the O radical on the N-vacancy is calculated to be -9.30~eV, while for the B-vacancy it is only -4.54~eV. 
However, reconstruction of the V$_\text{B}$:O by switching positions of the adjacent N and O atoms leads to a much more favourable configuration (V$_\text{B}$:O(R) in Fig.~\ref{fig: Figure 6}b) with the formation of a B--O--B unit.
In the case of double vacancies (Fig.~\ref{fig: Figure 6}c), the B--O--B configuration is energetically 3.7~eV more favourable than N--O--N. Similar trends are observed for the V$_9$ defects (Fig.~\ref{fig: Figure 6}d): binding of oxygen is stronger on the B-terminated edges.
Also, similar to the reconstruction of the V$_\text{B}$:O defect, rotation of the N--O bond in V$_9^\text{N}$:N--O--N structure leads to a lower-energy configuration, although the energy gain is only 0.46~eV. 

\begin{figure}[t]
    \centering
    \includegraphics{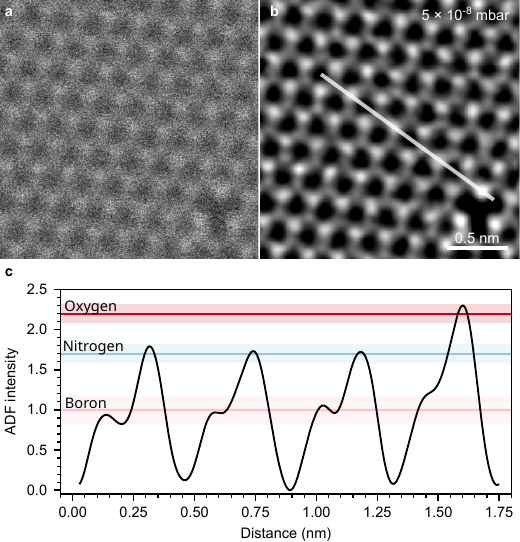}
    \caption{{\bf Oxygen atoms around a single B vacancy.} (a) STEM-MAADF image of hBN at a pressure of $5.7\times10^{-8}$~mbar, (b) the same image with a double-Gaussian filter applied and an overlaid line profile (top-left to bottom-right), (c) intensity of the line profile. Intensity of individual atom corresponds to $Z^{1.64}$. The intensity corresponding to boron has here been normalized to one, and the relative intensities corresponding to nitrogen and oxygen have also been highlighted on the plot.} 
    \label{fig: Figure 7}
\end{figure}

These results show that binding of O radicals to boron atoms is energetically much more favourable than binding to nitrogen atoms, with the B--O--B configuration being the most stable. 
We have verified that this holds true for at least a double vacancy, also for $\pm1$ and $\pm2$ charge states (Supplementary Table 2~\cite{supp2025}).
Interestingly, this has been previously seen in STEM images~\cite{li_prolonged_2023}, and we have also observed the B--O--B configuration in our data. Fig.~\ref{fig: Figure 7} shows the beginning of a pore formation under the electron beam in oxygen pressure of $5\times10^{-8}$~mbar. At first glance, we see bright atoms at the pore edges in the raw image (Fig.~\ref{fig: Figure 7}a), which are more evident in a double-Gaussian filtered image (Fig.~\ref{fig: Figure 7}b). A line profile~(Fig.~\ref{fig: Figure 7}b) and the corresponding intensity graph are shown in Fig.~\ref{fig: Figure 7}c. The values have been normalized against boron so that their ADF intensities are 1, and the other intensities have been calculated assuming $I \propto Z^{1.64}$~\cite{krivanek_atom-by-atom_2010}. Here we also see oxygen atoms sitting at nitrogen sites in B--O--B configuration, which is consistent with our DFT calculations.

To obtain a more realistic model of experimentally observed pores, we extended the supercell to $8\times8$ and used the projector-augmented wave DFT method with a linear combination of atomic orbitals (LCAO) basis set, better suited for larger systems~\cite{Mortensen2023} (Methods).
Two distinct triangular pore types were studied: one with edges terminated by nitrogen atoms and the other with edges terminated by boron atoms, both created by removing 17 atoms from the hBN lattice.
At the edges, both boron and nitrogen atoms are under-coordinated and seek stabilization by adsorbates, particularly atomic oxygen.
We then calculated the adsorption energies and performed nudged elastic band (NEB) simulations to estimate the energy barriers for the desorption of B–-O and N-–O species and found that adsorption of an oxygen radical on a boron atom releases 8.45~eV, while desorption of the B--O species requires overcoming a barrier of 6.78~eV.
By comparison, adsorption on a nitrogen atom releases 5.54~eV, and the desorption of N–-O requires overcoming a lower barrier of 2.97~eV (Supplementary Fig.~4--5~\cite{supp2025}).

Overall, these simulation results do not provide a complete picture of pore formation in hBN under electron irradiation in a low-pressure atmosphere. There are several possible explanations for this. Firstly, the experimental situation may be more complex than what is captured by the simulations, for example, due to the diffusion dynamics of oxygen radicals on the basal plane of hBN as they approach the pore edge, or structural changes caused by oxygen adsorption near the edge. Secondly, electron-beam-induced damage in semiconducting and insulating 2D materials has not been satisfactorily explained yet, even for pristine materials, despite recent efforts~\cite{kretschmer_formation_2020,speckmann_combined_2023, yoshimura2023quantum, bai2025excited}.

In conclusion, we have shown that pore formation in hBN is significantly affected by the presence of oxygen in the microscope column.
In UHV (at a pressure of $4\times10^{-10}$~mbar), pore growth in hBN under electron irradiation is significantly slower than at $4\times10^{-9}$~mbar.
The pore shape caused by direct electron irradiation in UHV is circular, but becomes triangular with nitrogen-terminated edges under a low-pressure oxygen atmosphere.
This result is confirmed across different sample sources and types, at various electron energies as well as focused scanning and defocused stationary beams.
We also demonstrated that molecular oxygen, by itself, does not lead to pore growth, as no changes are observed in the sample under an oxygen atmosphere when it is not irradiated, indicating that the electron beam is required to split molecular oxygen. It remains unclear why boron is removed exclusively via this mechanism, since, despite being energetically less favored, nitrogen atoms at the edge should also attract oxygen.
Nevertheless, we deduce from the experimental results that oxygen atoms must bind to boron atoms at or near the pore edge and can subsequently be removed, along with the oxygen, by the electron beam, leading to nitrogen-terminated triangular pores.
Overall, our results demonstrate that pore growth in hBN under electron irradiation at room temperature, within typical TEM column residual-vacuum pressures, results from an interplay between knock-on damage and chemical etching by oxygen.
Furthermore, the pore shape depends on the dominant process: knock-on damage forms round pores, while chemical etching favors triangular nitrogen-terminated pores.

\section{Methods}
\subsection{Sample preparation}
Commercially available hBN grown via chemical vapor deposition (CVD) on copper substrates (Graphene Supermarket, CVD1 and Sigma-Aldrich, CVD2) was transferred onto a Quantifoil Au TEM grid (hole diameter 2~$\mu$m or 0.6~$\mu$m, depending on the sample) by attaching the grid to the hBN/Cu using a drop of isopropyl alcohol (IPA), and subsequently etching away the substrate in approximately 0.04~M iron(III)-chloride (FeCl$_3$) solution, followed by multiple washing cycles in deionized water and IPA to remove residual etchant, as described in Ref.~\cite{ahmadpour_monazam_substitutional_2019}. 
The exfoliated hBN sample was produced using micromechanical exfoliation of commercially available bulk hBN (HQ Graphene).
A flake of hBN was carefully pushed against sticky tape to attach the top few layers to it.
After removing the flake, the tape was folded and peeled multiple times to exfoliate individual layers.
The material was transferred onto a SiO$_{2}$ chip by bringing it into contact with the sticky tape.
A Quantifoil grid was then thermally adhered onto a region of interest on the chip with IPA.
The SiO$_{2}$ chip was detached by chemical etching with a few drops of concentrated KOH solution, after which the flake was transferred onto the TEM grid.
The exfoliated sample had a minimum of three layers, as shown in Fig.~\ref{fig: Figure 1}. To reach the single layer, the electron beam was continuously scanned over the area until a monolayer region formed.  
Before insertion into the microscope, the samples were heated in vacuum at around 150°C for about 12~h to remove adsorbed water and weakly bound atmospheric contamination from the surface~\cite{tripathi_cleaning_2017}.

\subsection{In-situ electron microscopy}
The Nion UltraSTEM 100 microscope was used for imaging with a medium-angle annular dark-field (MAADF) detector with a semi-angular range of 60--200~mrad.
The convergence semi-angle was 35~mrad.
The beam current at the 60~kV alignment was around 110--140~pA, and at 80~kV it was around 30--40~pA.
The direct-electron detector Dectris ARINA was used to measure the electron dose rate at 80~kV~\cite{susi_quantifying_2025}.
For 60~kV experiments carried out before the detector was installed, the current was estimated as described in Ref.~\cite{speckmann_combined_2023}.
Most data was recorded as a stack of images to analyze the rate of pore growth.

The Vienna microscope has been modified to add extra pumping in the microscope column~\cite{hotz_ultra-high_2016} and a leak valve that allows introducing gases into the objective area during imaging.
Additionally, it is part of an interconnected system of vacuum instruments, allowing samples to be kept in ultra-high vacuum between experiments~\cite{mangler2022materials}.
This also means that the microscope column is not exposed to the atmosphere during sample exchange, maintaining a base pressure in the microscope at the sample is around $10^{-10}$~mbar (as measured by the objective area gauge), which is around three orders of magnitude lower than typical microscopes.
An equilibrium between leaking and pumping can be achieved to create a desired atmosphere up to $4\times 10^{-6}$~mbar during atomic-resolution imaging~\cite{leuthner2019scanning}.
The effect of electron irradiation was measured at a partial pressure range of $10^{-10}$--$10^{-8}$~mbar for oxygen and at a nitrogen partial pressure of $10^{-8}$~mbar.
All pressures reported here are readings of the objective area gauge.
In calculations of the impingement rates, a factor of 10 is included (as described in the text) to account for the estimated difference between the actual pressure at the sample and the reading of the gauge established in our previous work~\cite{leuthner2019scanning}.
All scale bars on microscopy images correspond to calibrated values based on the lattice constant of hBN.

\subsection{Data analysis}
To calculate the rate of pore growth, each image from the stack was analyzed.
Pores smaller than 0.5~nm$^2$, containing foreign atoms or merging during the measurement, were excluded from the analysis.
The area of all included pores was calculated and summed to obtain the pore area per image.
The pores in the images were defined using a threshold, with dark pixels contoured to obtain all pixels in a single pore.
The nm/pixel values of the images were calibrated based on the Fast Fourier Transform (FFT)~\cite{jacob_fourier_scale_calibration} and the lattice constant of hBN, which were then multiplied by the number of pixels in each pore to get their area.
Jupyter notebooks were used to analyze the data, and have been made available with the original data~\cite{javed2024}.

\subsection{Density functional theory simulations}
Total-energy DFT simulations for small vacancy structures were performed using the VASP package~\cite{Kresse1993,Kresse1996}. We used the PBE exchange-correlation functional~\cite{PBE} together with the D3 dispersion correction~\cite{Grimme2011}. hBN layers were modelled using hexagonal supercells consisting of 98 atoms with a vacuum layer of 15\,\AA. The plane-wave energy cutoff was set to 450~eV, and the Brillouin zone was sampled using a 3$\times$3$\times$1 $\mathbf{k}$-point mesh.
Structural relaxation was performed using the conjugate-gradient algorithm with the convergence criterion for the force of 0.01\,eV/\AA. In all calculations, spin polarization has been taken into account. Adsorption energies were calculated as ($E_\mathrm{adsorb}=E_\mathrm{hBN+O}-E_\mathrm{hBN}-E_\mathrm{O}$), where $E_\mathrm{hBN+O}$ is the total energy of the defective hBN with the adsorbed oxygen atom, $E_\mathrm{hBN}$ and $E_\mathrm{O}$ are the total energies of the defective hBN layer and a separate oxygen atom, respectively.

\subsection{Nudged elastic band simulations}
To perform nudged elastic band (NEB) simulations, we employed DFT using the GPAW software package~\cite{Mortensen2023}. Here, the pristine hBN model consisted of an 8$\times$8 supercell containing 128 atoms, whereas the defective supercell, featuring a pore created by removing 17 atoms, contained 111 atoms.
The supercell was sufficiently large to allow for $\mathbf{k}$-space sampling using only the $\Gamma$ point.
For all boron and nitrogen atoms, we used a double-zeta polarized (dzp) localized atomic orbital (LCAO) basis set, and a grid spacing of 0.2\,\AA.
NEB paths were calculated with ten intermediate images to explore the adsorption and dissociation pathways.
Also these calculations were spin polarized.

\section{Data Availability}
All the data used for producing the results are available via the University of Vienna's PHAIDRA repository at 
https://phaidra.univie.ac.at/o:2144401~\cite{javed2024}.

\bibliographystyle{apsrev4-1}
\bibliography{references}

\section{Funding Statement}
This research was funded in part by the Austrian Science Fund (FWF) [10.55776/COE5 and 10.55776/P36264].
For open-access purposes, the author has applied a CC-BY public copyright license to any author-accepted manuscript version arising from this submission.

\section{Author contributions}
UJ and JK conceived the study.
UJ, CM, BMM, ML, CK, and MP conducted the experimental work.
UJ, MP, ML, and JK performed the data analysis.
CK, BMM, DL, and NR prepared the samples.
VZ and AM performed the \textit{ab initio} calculations.
UJ, ML, and JK wrote the manuscript, with contributions from all authors.
TS and JK supervised the study.

\section{Competing Interests}
The authors declare no competing interests.

\include{supplement.tex}

\end{document}

%% file: supplement.tex
\onecolumngrid

\begin{center}
    \large {Supplementary Information}\\[1em]
    \large \textbf{Origin of circular and triangular pores in electron-irradiated hexagonal boron nitride}\\
    \large
    Umair Javed$^{1,2}$, Manuel Längle$^{1}$, Vladimír Zobač$^{1}$, Alexander Markevich$^{1}$, Barbara Maria Mayer$^{1,3}$ Clara Kofler$^{1,2}$, Martin Paul$^{1}$, Darwin Lorber$^{1}$, Nandhini Ravindran$^{1,2}$, Clemens Mangler$^{1}$, Toma Susi$^{1}$, Jani Kotakoski$^{1}$\\[1em]
    \normalsize
    $^{1}$ University of Vienna, Faculty of Physics, Boltzmanngasse 5, 1090 Vienna, Austria\\
    $^{2}$ University of Vienna, Vienna Doctoral School in Physics, Boltzmanngasse 5, 1090 Vienna, Austria\\
    $^{3}$ Uppsala University, Faculty of Physics, Box 516, SE-751 20, Sweden\\[1em]
\end{center}
\setcounter{figure}{0}  
\begin{figure}[htbp]
    \centering
    \includegraphics{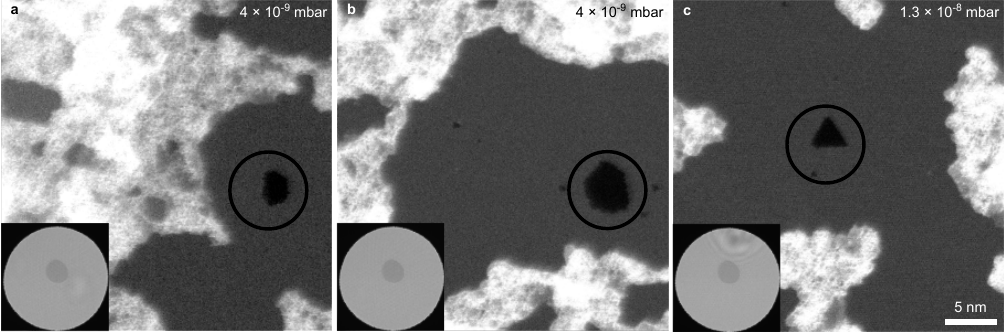}
    \caption{Pore formation in parallel-beam (CCD) mode. Different areas of hBN were irradiated for roughly 4~minutes with 60~kV electrons, a beam current of ca. 118~nA, with a defocus of 125~nm, and fluence of ca. $4.9\times10^{9}$~e/nm$^2$. (a-b) Oxygen partial pressure of $4\times10^{-9}$~mbar. (c) Oxygen partial pressure of $1.3\times10^{-8}$~mbar. The black circle roughly outlines the irradiated area, the insets in the lower-left corners show how the sample appeared on the CCD camera, and the scale bar corresponds to all the images.
    Even in parallel-beam mode, in an oxygen partial pressure in the range of low $10^{-8}$~mbar, the pores that emerge are not triangular, but when the pressure is increased to the $10^{-8}$~mbar range, the pores that are formed are triangular.}
    \label{fig.s1}
\end{figure}

\begin{table}[htbp]
    \caption{The shapes of pores formed in parallel-beam mode (CCD), in different oxygen pressures, and at different defoci. In these data, the pressure gauge reading at the objective area is used without any additional factor arising from the sample stage geometry (see the main text).}
    \centering
    \begin{tabular}{c c c c c}
        \hline
        Pressure (mbar) & Defocus (nm) & $J\,(\mathrm{O}_{2}$s$^{-1}$nm$^{-2}$) & Dose (e$^{-}$s$^{-1}$nm$^{-2}$) & Shape\\  
        \hline
        $4\times10^{-9}$ & 125 & 2 & $1.7\times10^{7}$ & Non-triangle \\
        $4\times10^{-9}$ & 125 & 2 & $1.7\times10^{7}$ & Non-triangle \\  
        $4\times10^{-9}$ & 125 & 2 & $1.7\times10^{7}$ & Non-triangle \\  
        $4\times10^{-9}$ & 125 & 2 & $1.7\times10^{7}$ & Non-triangle \\
        $4\times10^{-9}$ & 125 & 2 & $1.7\times10^{7}$ & Non-triangle \\
        $1.3\times10^{-8}$ & 125 & 6 & $1.7\times10^{7}$ & Triangle \\
        $4\times10^{-9}$ & 500 & 2 & $1\times10^{6}$ & Triangle \\
        $4\times10^{-9}$ & 500 & 2 & $1\times10^{6}$ & Triangle \\
        $4\times10^{-9}$ & 500 & 2 & $1\times10^{6}$ & Triangle \\
        $1.3\times10^{-8}$ & 500 & 6 & $1\times10^{6}$ & Triangle \\
        \hline
    \end{tabular}
    \label{tab: Table 1}
\end{table}
\clearpage
\begin{figure}[htbp]
    \centering
    \includegraphics{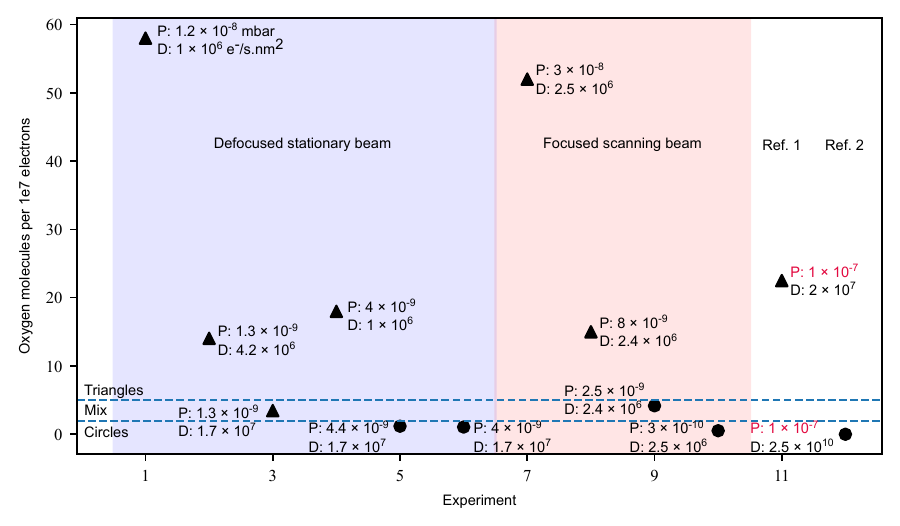}
    \caption{Plot of the ratio of oxygen molecules impinging per 10 million electrons, and the pore shape in hBN at those conditions. The horizontal axis shows different types of experiments, and the markers indicate whether a triangle or a circle corresponds to the pore shape. For comparison with studies in the literature, we used dose-rate values from Meyer et al.~\cite{meyer2009selective} and Keneipp et al.~\cite{keneipp_nanoscale_2024}, and assumed a pressure of $1\times10^{-7}$~mbar. In all data presented here, the pressure gauge reading at the objective area is used without any additional factor arising from the sample stage geometry (see the main text).}
    \label{fig.s2}
\end{figure}

\begin{table}[htbp]
    \caption{Rate of oxygen molecules impinging on the surface of the sample within a 16~nm$^2$ field of view, or an atomic site per frame, at each partial pressure (site here refers to a lattice site in a pristine hBN structure). In these data, a factor of 10 has been added to the impingement rate to account for the sample stage geometry (see the main text).}
    \centering
    \begin{tabular}{c c c c}
        \hline
        Pressure ($10^{-8}$ mbar) & $J\, (10^{18} $m$^{-2}$s$^{-1})$ & $J_\mathrm{FOV}$ & $J_\mathrm{Site}$ \\  
        \hline
        0.4 & 0.12 & 527 & 0.05 \\
        0.8 & 0.21 & 949 & 0.1 \\  
        1.7 & 0.46 & 1992 & 0.2 \\  
        2.5 & 0.68 & 2929 & 0.3 \\
        3.3 & 0.89 & 3867 & 0.4 \\
        4.1 & 1.1 & 4804 & 0.5 \\
        4.9 & 1.3 & 5742 & 0.6 \\
        5.7 & 1.5 & 6679 & 0.7 \\
        6.5 & 1.7 & 7617 & 0.8 \\
        7.3 & 1.9 & 8554 & 0.9 \\
        8.1 & 2.1 & 9492 & 1.0 \\
        \hline
    \end{tabular}
    \label{tab: Table 2}
\end{table}

\clearpage
\begin{table}[htbp]
    \caption{Calculated total-energy difference between B--O--B and N--O--N configurations for different charge states of the double-vacancy defect in hBN with an oxygen atom (see the main article for the atomic configurations).}
    \centering
    \begin{tabular}{c c}
        \hline
        Charge ($e^-$) & $\Delta E$ (eV) \\  
        \hline
        -2 & -7.20 \\
        -1 & -6.61 \\
         0 & -3.70 \\
        +1 & -3.10 \\
        +2 & -3.07 \\
        \hline
    \end{tabular}
    \label{tab: Table 3}
\end{table}

\begin{figure}[htbp]
    \centering
    \includegraphics{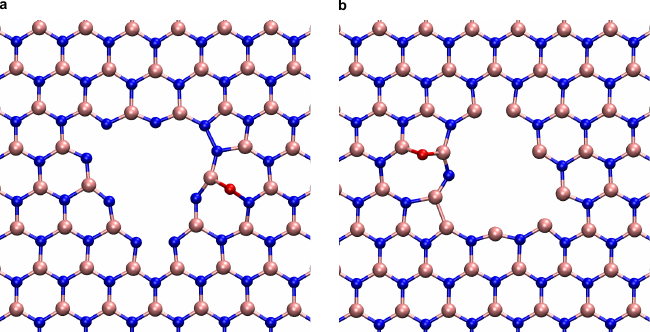}
    \caption{Adsorption of oxygen on hBN near the edges of V$_9$ defects. (a) In the case of V$_9^\text{N}$, oxygen radical adsorption on the B--N bond close to the edge leads to breaking of the bond (Fig.~6b), with the distance between B and N atoms becoming $\sim$2.5\,\AA. The adsorption energy is calculated to be -3.94\,eV, which is significantly lower than that on the pristine surface. (b) Adsorption of O near the B-terminated edge leads to even more drastic changes in the bonding (Fig.~6c): the N atom is pushed towards the edge and a B--O--B bridge configuration is formed, resulting in the adsorption energy of -6.10\,eV.} 
    \label{fig: s3}
\end{figure}
\clearpage
\begin{figure}[htbp]
    \centering
    \includegraphics[width=80mm]{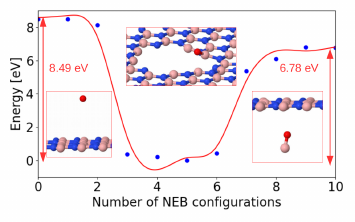}
    \caption{Energy profile of the oxygen adsorption and desorption process on a boron-terminated V$_9^\mathrm{B}$ edge obtained from NEB simulations. The inset panels depict the initial, intermediate, and final configurations along the reaction path. The red curve serves as a visual guide, generated by spline interpolation of the discrete NEB energy points.} 
    \label{fig: s4}
\end{figure}
\begin{figure}[htbp]
    \centering
	\includegraphics[width=80mm]{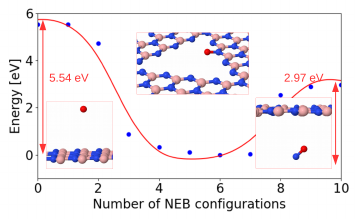}
    \caption{Energy profile of the oxygen adsorption and desorption process on a nitrogen-terminated V$_9^\mathrm{N}$ edge obtained from NEB simulations. The inset panels depict the initial, intermediate, and final configurations along the reaction path. The red curve serves as a visual guide, generated by spline interpolation of the discrete NEB energy points.} 
    \label{fig: s5}
\end{figure}
\clearpage